\documentclass[12pt,preprint]{aastex}

\slugcomment{Not to appear in Nonlearned J., 45.}

\shorttitle{The open cluster NGC~6520}
\shortauthors{Carraro et al.}

\begin{document}

\title{The open cluster NGC~6520 and the nearby dark
molecular cloud Barnard~86}

\author{Giovanni Carraro\altaffilmark{a,b}, Ren\'e A. M\'endez, Jorge May
and Diego Mardones}
\affil{Departamento de Astronom\'ia, Universidad de Chile,
    Casilla 36-D, Santiago, Chile}
\email{gcarraro,rmendez,jorge,mardones@das.uchile.cl}

\altaffiltext{a}{Yale University, Astronomy Department, New Haven, CT 06511, 
USA}
\altaffiltext{b}{On leave from: Dipartimento di Astronomia, Universit\`a
di Padova, Vicolo Osservatorio 5, I-35122 Padova, Italy}

\begin{abstract}
Wide field BVI photometry and $^{12}$CO(1$\rightarrow$0) observations are presented
in the region of the open cluster NGC~6520 and the dark molecular cloud
Barnard~86. From the analysis of the optical data we find that the cluster is 
rather compact,
with a radius of 1.0$\pm$0.5 arcmin, smaller than previous estimates.
The cluster age is 150$\pm$50 Myr and the reddening E$_{B-V}$=0.42$\pm$0.10. The distance
from the Sun is estimated to be  1900$\pm$100 pc, and it is larger than previous estimates.
We finally derive basic properties of the dark nebula Barnard~86 
on the assumption that it lies at the same distance of the cluster.
\end{abstract}


\section{Introduction}
NGC~6520 (C1800-279, $\alpha=18^{\rm h}~03^{\rm m}.4$,
$\delta=-27^{\circ}
54^{\prime}.0$, $l=2^{\circ}.87$, $b=-2^{\circ}.85$, J2000.) 
is a compact moderate age open cluster located in the Sagittarius
constellation only 4 degrees eastward of the Galactic Center direction.
The region is very crowded (see Fig.~1), and harbors also the dark molecular
cloud Barnard~86 ($l=2^{\circ}.85$, $b=-2^{\circ}.75$, \citealt{bar27}), a few arcmin westward of the cluster. The cloud is a very
prominent feature, but other dust lanes are present across the field.
In particular, the feature extends toward the cluster, and it is readily visible southward. 
Moreover, by inspecting more closely Fig.~1, or sky maps, one has the clear
impression that the dark nebula encompasses the whole cluster.\\ 
This implies that across the region we expect the reddening to vary significantly,
although within the small cluster radius 
variable reddening should not be a real issue.\\
It seems quite natural to look for a possible relationship between
the cluster and the cloud, due to their proximity.
Their mutual relationship can in fact turn out to be quite important for our understanding of the 
timing and outcomes of star clusters formation process.\\
To address this issue, we obtained Wide Field optical photometry in the BVI filters,
and $^{12}$CO(1$\rightarrow$0) data, which are presented and discussed in this paper.
The aim is to derive estimates of the cluster and cloud distances from the Sun,
which is the basic step to clarify their possible relationship.\\

\noindent
The paper is organized as follows. Section~2 illustrates the data acquisition details
and the data reduction procedures. Section~3 is dedicated to the analysis of the 
optical data and the derivation of NGC~6520 fundamental parameters. Section~4
deals with Barnard~86 physical parameters determination, while Section~5
highlights the paper results and proposes future lines
of investigation.

\section{Observations and Data Reduction}
This Section illustrates the data acquisition and the employed
reduction techniques.

\subsection{Optical photometry}
BVI photometry of two overlapping fields 
in the region of NGC~6520 and Barnard~86
was taken at the Cerro Tololo Inter-American Observatory (CTIO) 0.90m telescope
on the nights of 24 and 25 June 1999.
The pixel scale of the 2048 $\times$ 2046 Tek2k $\#$3 CCD
is 0.396$^{\prime\prime}$, allowing to observe a field of 13.5 $\times 13.5$ 
squared arcmin in the sky.
The total covered area amounts to $20.0 \times 13.5$ squared arcmin.
The two nights were photometric with an average seeing of 1.3 arcsec.
We took several short (5 secs), medium (300 secs) and long (600 secs)
exposures
in all the filters to avoid saturation of the brightest stars.
The data have been reduced with the
IRAF\footnote{IRAF is distributed by NOAO, which are operated by AURA under
cooperative agreement with the NSF.}
packages CCDRED, DAOPHOT, ALLSTAR and PHOTCAL using the point spread function (PSF) 
method \citep{ste87}.
Calibration was secured through the observation of \citet{lan92}
standard fields  G~26, PG~1047, PG~1323, PG~1633, PG~1657, 
SA~110, and SA~112 for a grand total of 90 standard stars.
The calibration equations turned out of be of the form:\\

\noindent
$ b = B + (2.778\pm0.007) + (0.25\pm0.01) * X - (0.092\pm0.007) \times (B-V)$ \\
$ v = V + (2.595\pm0.008) + (0.16\pm0.01) * X + (0.030\pm0.007) \times (B-V)$ \\
$ v = V + (2.595\pm0.008) + (0.16\pm0.01) * X + (0.026\pm0.006) \times (V-I)$ \\
$ i = I + (3.554\pm0.005) + (0.08\pm0.01) * X + (0.029\pm0.004) \times (V-I)$ ,\\

\noindent
and the final {\it r.m.s} of the calibration turn out to be 0.03 for all the 
pass-bands.\\
Photometric errors have been estimated following \citet{pat01}.\
It turns out that stars brighter than
$V \approx 20$ mag have
internal (ALLSTAR output) photometric errors lower
than 0.15~mag in magnitude and lower than 0.21~mag in colour.
\noindent
The final photometric data (coordinates,
B, V and I magnitudes and errors)
consist of 50,000 stars will be made
available in electronic form at the
WEBDA\footnote{http://obswww.unige.ch/webda/navigation.html} site
maintained by J.-C. Mermilliod.\\
We finally estimated the completeness of our samples in the $V$ and $I$ filters. 
The completeness corrections have been determined by standard artificial-star
experiments on our data (see \citealt{gca05}).
Basically we run several times the ADDSTAR task, adding 15~\% of the number
of detected stars and by counting the number of stars with the ALLSTAR task
after performing all the steps and by using the same parameters as for the original
images. In each run, a new set of artificial stars, following a similar
magnitude distribution, was added at random positions. A comparison between the
amount of added and detected stars, taking also into account that not all the
stars are detected in both $V$ and $I$ filters, yield a completeness level
of 100$\%$ down to V = 16, and larger than 60$\%$ down to V=20. At fainter magnitudes,
the completeness level falls below 50$\%$, and therefore we are considering  V = 20 
as our  limiting magnitude for the purpose of deriving the cluster mass (but see
also Sect.~3.1).

\subsection{CO observations}
The observations were carried out in September 1998 
with the Columbia U.-U. de Chile
Millimeter-wave Telescope (\citealt{coh83},  \citealt{bro88}) located
at Cerro Tololo Interamerican Observatory, Chile. The telescope is a
1.2-m Cassegrain with a beam-width of 8.8\arcmin\,(FWHM) at 115 GHz,
the frequency of the CO($J=1\rightarrow0$) transition. It was equipped
with a super-heterodyne receiver with a SSB noise temperature of 380 K.
The first stage of the receiver consisted of a Schottky barrier
diode mixer and a GaAs field-effect transistor amplifier cooled to
77 K by liquid nitrogen.

The spectrometer was a 256-channel filter bank of standard NRAO
design. Each filter, 100 kHz wide, provided a velocity resolution
of 0.26 km s$^{-1}$ at 115 GHz and a coverage of 66 km s$^{-1}$.
The integration time for each position varied between 10 and 12
minutes, depending on source altitude and atmospheric opacity.
Spectra were intensity calibrated individually against a blackbody
reference by the standard chopper-wheel method (e.g \citealt{kut81}
and references therein), yielding a temperature scale $T_{a}^{*}$
corrected for atmospheric attenuation, resistive losses, and rearward
spillover and scattering. Sampling interval of 3.75\arcmin
(0$^o$.0625) was used, almost half beam-width, to optimize the angular
resolution of the instrument. Position switching was used for all the
observations with equal amounts of time spent on the source and on the
reference positions. This radiotelescope has a main beam
efficiency of 0.82 (\citealt{bro88}).

\section{Study of the Open Cluster NGC~6520}
In this section we use the optical data to derive NGC~6520
basic parameters, namely radius, reddening, distance,
and age by means of star counts and isochrone fitting in the Color 
Magnitude Diagrams (CMDs).\\ 
The cluster was already studied in the past.
The most recent study is by 
\citet{kjf91}, who  obtained CCD UBV photometry of about 300 stars
in a $4.0 \times 2.6$ arcmin area, and 
found that NGC~6520 is a 190 million years old cluster
located 1.60 kpc from the Sun with a reddening E$_{B-V}$ =
0.43. These results basically agree with previous investigation
\citep{sve66}.
We observed 50,000 stars down to V $\approx$ 21. We compared our photometry
with \citet{kjf91} one, and generally found a very good agreement at the level
of 0.01 mag, but for the 
the brightest stars (V $\leq$ 11.4), where our stars are   brighter than 
\citet{kjf91} ones by 0.2-0.3 mag. We believe that this is due to saturation
problems in \citet{kjf91} photometry.\\

\subsection{Star counts and cluster size}
The aim of this section is to obtain the surface density distribution
of NGC 6520, and derive the cluster size in the
magnitude space by means of star counts. The cluster radius is indeed
one of the most important cluster parameters, useful (together with
cluster mass) for a determination of cluster dynamical parameters.
Star counts allow us to determine statistical properties of clusters (as
visible star condensations) with respect to the surrounding stellar
background.\\
By inspecting Fig.~1, NGC 6520 appears as a concentration of bright stars
in a region of about 1-2 arcmin.
In order to derive the radial stellar surface density we first seek for the
highest peak in the stellar density to find out the cluster center.
The adopted center is placed at $\alpha = 18:03:24.0$; $\delta =
-27:53:18.0$, as given by \citet{dia02}.
Then, the radial density profile is constructed by
performing star counts inside increasing concentric annuli $0\farcm5$
wide, around the cluster center and then by dividing by their
respective surfaces. This is done as a function of apparent magnitude,
and compared with the mean density of the surrounding Galactic field
in the same brightness interval. The contribution of the field
has been estimated through star counts in the region outside 6 arcmin
from the cluster center. Poisson standard deviations have also been
computed and normalized to the area of each ring as a function of magnitude,
both for the cluster and for the field.\\
The result is shown in Fig.~2, where
one readily sees that NGC~6520 significantly emerges from the mean
field above V$\approx$18. At fainter magnitudes the cluster gets
confused with the Galactic disk population. Based on the radial density
profiles in Fig.~1, we find that stars brighter than V=18
provide a cluster radius smaller than 2 arcmin.
We adopt as a final estimate
of the radius the values $1.0\pm0.5$ arcmin. This is somewhat
smaller than the estimate of 2.5 arcmin reported by \citet{dia02},
which was simply based on visual inspection.  We shall adopt this values of
the cluster radius throughout this work. \\
We stress however that this
radius is not the limiting radius of the cluster, but the distance
form the cluster center at which  the cluster population starts to be confused
with the field population.

\subsection{Analysis of the CMDs}
The BVI photometry allow us to build up CMDs in the V vs (B-V) and V vs (V-I)
plane. They are shown in Fig.~3 panels, where we consider only the stars located 
within 1 arcmin from the cluster center. The Main Sequence (MS) extends from V = 11 to V = 20,
although at V = 18 the contamination by foreground stars starts to be significant.
The MS has a width of  about 0.2-0.3 mag, a bit 
larger than expected from simply photometric errors (see Sect~2). There might be some differential reddening
across the cluster area, but we believe it is not very important.
Unfortunately we do not have U photometry, but from the two color (U-B) vs (B-V) diagram 
by \citet{kjf91}, one readily sees that the amount of differential reddening does not exceed
0.10 mag. Therefore, 
differential reddening does not play an important role in a cluster
that compact, and the MS width might be the result of the combination
of the two effects, plus some probable interlopers and/or binary stars.
The detailed shape of the Turn Off (TO) region
is much better defined than in \citet{kjf91}, and deserves more attention, since the scatter
of stars there is much larger than in the MS. The MS seems to stop at V = 12.5 and (B-V) = 0.25
( (V-I) = 0.35) , and the brighter
stars clearly are on the act of leaving it, since they lie red-ward of the MS. 
We are keen to believe that the few stars above the TO, brighter than V = 12, 
are field stars or blue stragglers, which are frequent in open star clusters.
The global shape of the CMD is that of an open cluster of moderate age, like
NGC~6204 (\citealt{car04}) or NGC~2287 (\citealt{har93}).

\subsection{NGC~6520 basic parameters}
The fundamental parameters of NGC~6520 have already been derived in the past.
Both \citet{sve66} and \citet{kjf91} agree that the cluster possesses a
heliocentric distance of about 1600 pc and a reddening E$_{B-V}$=0.42.
As for the age, there are some contradictory findings in the literature.
In fact, while \citet{sve66} reports an age of 800 million years, \citet{kjf91}
suggest a much younger age of 190 million years.\\
This is clearly due to the scatter in the upper part of the MS, which makes it difficult
to recognize the Turn Off Point (TO).
Before analysis our data we made use of the \citet{kjf91} $UBV$ photometry to select
photometric members basing on the $Q$ method (\citealt{car02}), and from 
the reddening corrected CMDs (see Fig.~4) we derived an absolute distance modulus
$(m-M)_o=11.40$ by fitting an empirical  
Zero Age Main Sequence (ZAMS) from \citet{sch82}. 
Although the sequence is vertical and there might be problem in the brightest
stars magnitude (see Sect.~3),
the fit is quite good in both the CMDs and provides a distance 1900$\pm$100 pc.\\

\noindent
Additionally,
we derive NGC~6520 fundamental parameters by directly
comparing the CMD with theoretical isochrones
from the Padova group (\citealt{gir00}). 
The results of the fit are presented
in Fig.~5 and Fig.~6. 
In the left panel of Fig~5 we plot the V vs (B-V) CMD for all the stars within 1.5 arcmin from
the cluster center. Over-imposed is a ZAMS from \citep{sch82} 
(dashed line) shifted by (m-M)$_V$ = 12.75$\pm0.10$ and E$_{B-V}$=0.42$\pm$0.10
(error by eyes), 
which fits very well the bulk of the cluster MS. Together with the ZAMS we over-imposed two
solar metallicity isochones for the age of 100 (dashed line) and 200 (dotted line) million years
from \citet{gir00} shifted by 
(m-M)$_V$ = 12.65 and E$_{B-V}$=0.40 and (m-M)$_V$ = 12.75 and E$_{B-V}$=0.42,
respectively.
This fit provides an estimate of the cluster distance from the Sun,
which turns out to be 1950 pc. This estimate is about 20\% larger than the previous one
by \citet{kjf91}, and we believe this is due to the saturation of bright stars in 
\citet{kjf91} which prevented to define clearly the upper part of the MS.\\
However, on the base of this CMD one cannot exclude a younger age due to the presence
of several bright stars along the ZAMS.
To better understand the nature of these stars, we isolate in the right panel 
of the same figure the stars
within 0.5 arcmin from
the cluster center, to minimize the effect of field stars contamination.
The fit is still very good, and confirms previous findings about the distance and reddening,
suggesting that the bright stars are just interlopers.\\

\noindent
As for the age, a close scrutiny of the TO region clearly favors an age around 200
million years. This is confirmed by the absolute magnitudes and colours of the brightest
photometric members. With a TO located at V $\approx$ 12.5-13, we expect that un-evolved stars
still in the MS have M$_V$ in the range  -0.25 to 0.25, and therefore a spectral type around 
B5-B8, which implies an age around 150 million year (\citealt{gir00}).\\

\noindent
Finally, in Fig.~6 we achieve the same fit both in the B vs (B-V) (left panel)
and in the V vs (V-I) diagram (right panel).
Here we 
provide again a reasonable fit to the data for the same distance modulus, and a reddening
E$_{V-I}$=0.55. The ratio $\frac{E_{V-I}}{E_{B-V}}$ turns out to be 1.30, not far from
the widely used 1.24 value from \citet{dea78}.\\

\noindent
In conclusion, we estimate the age of NGC~6520 to be 150$\pm$50 million years, and provide
a distance of about 1900$\pm$100 pc, $20\%$ larger than previous estimates.

\subsection{NGC 6520 mass}
In order to derive the cluster mass we firstly derive the cluster
luminosity function (LF) and mass function (MF) , in the same way as in \citet{gca05}.
Briefly, we considered only the stars within 1.5 arcmin from the cluster nominal center,
and brighter than V = 18 (see Sect.~3.1),
and statistically derived the cluster population by subtracting the field
population from a same area region outside 6 arcmin from the cluster nominal center.
The subtraction is made bin by bin, with bin size of 0.5 mag, and any bin
population has been corrected for the corresponding incompleteness.
The MF turns out to be a power law with slope $\alpha=2.4\pm0.3$ and therefore
compatible with the standard \citet{sal55} MF. By integrating the MF over the V mag.
range 11 to 18, we estimate that the cluster mass is 364$\pm54$ $M_{\odot}$.

\section{The dark Cloud Barnard~86}
In this section we derive the physical parameters of the molecular cloud associated with
Barnard 86 obtained from a Gaussian fit to its composite spectrum,
i.e., the sum of all spectra across the cloud's projected surface.\\
The quality of the spectra is shown is Fig.~7, together with the best Gaussian fit.
The figure shows the average of the two brightest CO 1$\rightarrow$0 spectra.
The cloud is
clearly detected in at least 9 positions within a 30 position map.
The {\it rms} noise of the spectrum shown is 0.15 K.  The intensity scale is
in Antenna temperature, uncorrected by the main beam efficiency.
This is a 9 sigma detection at the peak, or 15 sigma detection of the
integrated intensity.
The only effect of combining the central two spectra is to reduce the
noise by a factor 1.41, but the detection is strong in both spectra.

\noindent
\subsection{Cloud distance}
We derive the cloud properties by assuming that the nebula lies at the same distance of the cluster
NGC~6520. This is naturally a crude assumption, which is mainly motivated by the appearance
of Fig.~1. The center of the dark nebula lies 6 arcmin from the cluster center, but the two
objects do not appear clearly detached. The whole cluster seems to be surrounded by some nebulosity,
which gets denser in the southern regions, and which seems to form a bridge with the main body
of the cloud northwest of the cluster. This is also confirmed by the appearance of the whole region
in the Southern H-Alpha Sky Survey Atlas (\citealt{gau01}), where one can clearly see that the nebula
embraces the cluster.\\
Obviously,  this assumption must be verified in some
more quantitative way, which is not possible with the present data. \\
\noindent
The only way we envisaged was to count the number of stars in front of the cloud,
and compare this number with the number of stars expected to lie within 1, 2 and 3 kpc from the Sun
in the direction of the cloud, as derived from a galactic model (\citealt{men96}, \citealt{men98}).
From our photometry we can count in front of the dark cloud 15 stars
in the magnitude range $10 \leq V \leq 19$. In the same magnitude range, the Galactic model
predicts 3,046, 17,138 and 41,360 stars per square degree within 1, 2 and 3 kpc from the Sun.
By assuming an area of the dark cloud of 4 squared armin, we expect
from the model about 3.5, 19 and 46 stars in the same magnitude range in front of the nebula.
Although model-dependent, this results brings some support to our assumption of common distance
for the cluster and the dark cloud.

\noindent
\subsection{Results}
The peak of the CO emission turns out to be located at l=2.875$\degr$ and b=-2.8125$\degr$.
The radial velocity $V_{LSR}$ 
and the linewidth $\Delta V_{obs}$ 
result to be  11.28$\pm$ 0.25 and 2.07$\pm$0.25  km s$^{-1}$,
and corresponds to the peak and to the FWHM of the Gaussian fit of the composite
spectrum, respectively . 
The value of the velocity by the way is in nice agreement with the 11.6  km s$^{-1}$
estimate by \citet{cle88}.
Unfortunately, it is not possible to estimate the distance of the cloud from its radial
velocity by means of the Milky Way rotation curve, due to the particular cloud position,
located in the direction of the Galactic center.\\
\noindent
Adopting the 2 kpc distance, we estimate an effective 
radius of  5.36$\pm$0.56 pc. We define the effective radius of the cloud as
$\sqrt{A/\pi}$, where $A$ is the actual projected
area obtained from the angular extent of the cloud, measured
from the spatial maps.\\
\noindent

The luminosity of CO $L_{CO}$ was found to amount to 
5.1$\pm$0.2 $10^{-4}$ K km s$^{-1}$ pc$^{2}$, being
$L_{CO}$  the $^{12}$CO luminosity given by
 
\begin{equation}
     L_{CO} = d^{2} I_{CO} \,,
\end{equation}

\noindent
where $I_{CO}$ is the $^{12}$CO line intensity integrated
over all velocities and lines of sight within the boundaries
of the cloud, and $d$ is the heliocentric distance of the
cloud (2 kpc).

As for the mass,
this is computed from the $^{12}$CO luminosity, and results to be
$M_{CO} = 600\pm150 M_{\odot}$.
$M_{CO}$ was estimated directly from
its $^{12}$CO luminosity on the empirically based assumption
that the integrated CO line intensity is proportional to the
column density of H$_{2}$ along the line of sight (e.g.
\citealt{blo86} and \citealt{hun97}).
Thus, the masses were computed using the relation

\begin{equation}
     M_{CO} = w X L_{CO} \,,
\end{equation}

\noindent
where $w$ is the mean molecular weight per H$_{2}$ molecule,
$X$ is the constant ratio of H$_{2}$ column density to
integrated $^{12}$CO intensity, and $L_{CO}$ is the $^{12}$CO
luminosity. A detailed description on this kind of mass  calculation 
can be found in \citet{may97} and \citet{mur91}.\\
\noindent
It is also possible to derive an estimate of the cloud
virial mass.
The virial mass $M_{VT}$ was derived using the relation

\begin{equation}
M_{VT} = 126\, r_{c}\, (\Delta V_{obs})^{2} \,,
\end{equation}

\noindent
where $M_{VT}$ is in solar masses, $r_{c}$ is the effective
radius of the cloud, in pc, and $\Delta V_{obs}$ is the HPFW
of the cloud's composite spectrum, in km s$^{-1}$. This equation
assumes that : 1) the cloud is in virial equilibrium,
2) the cloud is spherical with a $r^{-2}$ density distribution,
where $r$ is the distance from its center,
3) the observed $^{12}$CO line width of the cloud is an accurate
measure of the net velocity dispersion of its internal mass
distribution, which is believed to be clumpy on many scales
Under these assumptions, the mass turns out be be
$M_{VT}$ = 3000 $M_{\odot}$.\\
Because of the large discrepancy between $M_{CO}$ and $M_{TV}$,
we can conclude that this molecular cloud is not in virial equilibrium,
and therefore its mass is about 600 $M_{\odot}$.

\section{Conclusions}
In this paper we have presented deep CCD BVI photometry and $^{12}$CO(1$\rightarrow$0)
in the region of the open cluster NGC~6520 and the dark molecular cloud
Barnard~86. We provided updated estimates of the cluster parameters,
suggesting that its heliocentric distance is 1900$\pm$100 pc, and the age is 150$\pm$50
million years.\\ 
\noindent
We then derive  Barnard~86 physical parameters on the assumption
that the cloud is at the same distance as the cluster.\\

\noindent
As we discussed, this assumption in the present study is only motivated by the 
particular distribution of the nebulosity in the region of the cloud and the cluster,
which seems to support some association of the two objects. 
In Fig.~8 we show a DSS map of the region where the cluster and the cloud are located,
and we overimposed the iso-intensity contour in $^{o}K \times Km sec^{-1}$.
The thick line is the Full Width Half Maximum (FWHM) of the Gaussian
fit, and clearly crosses the cluster. To us this might mean that the nebolosity
close to the cluster is related to Barnard~86, although better resolution
spectra are necessary to derive firmer conclusions.
We 
corroborated this hypothesis by comparing the number of stars we can see in front of the 
cloud with the predictions of a Galactic model, and it turned out that thei distance of 2 kpc is
quite reasonable.\\

\noindent
The age of the cluster (150$\pm$50 million years) seems to be 
incompatible with this assumption, since the mean lifetime
of a molecular cloud does not exceed a few time 10-million years (\citealt{bli80}).
This makes the hypothesis of the connection between the cloud and the cluster
puzzling.\\

\noindent
However, if we accept the common distance hypothesis, the scenario 
one might  envisage is that both NGC~6520 and
the dark cloud Barnard~86 share the same origin, the latter being the remain
of a star formation process finished about 150$\times 10^6$ years ago.\\
In this case, Barnard~86 may be an interesting cloud to study the possible
stability of gas condensations over a full Galactic rotation.
In fact, 
to our best knowlegde no other clouds are known to show evidences
of a possible relation with an open cluster that old.

\noindent
Clearly,
our assumption on the distance must be better constrained, and we propose
to derive for
instance the radial velocity of a few NGC~6520 member stars which should
be compared with the cloud velocity or the IR spectra of some bright stars
inside the nebula to measure the properties of the H$_{3}^{+}$ line
(\citealt{mcc03}).\\

\acknowledgments
We acknowledge fruitful discussions with S. Casassus and 
P. Caselli.
The work of GC is supported by {\it Fundaci\'on Andes}.
R.A.M., J.M., and D.M. acknowledge support from the
Chilean {\sl Centro de Astrof\'\i sica} FONDAP No. 15010003.
J.M. acknowledges partial support from FONDECYT through
grant 1010431.

\clearpage

\begin{figure}
\plotone{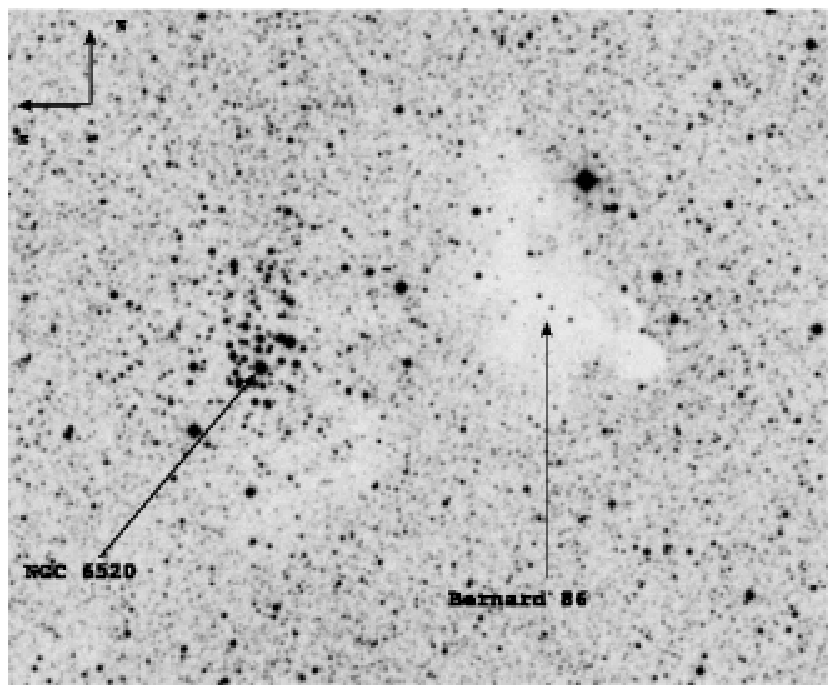}
\caption{A DSS map of the observed field in the direction of NGC 6520 and Barnard 86.
The size of the field is $20 \times 13.5$ arcmin.}
\end{figure}

\begin{figure}
\plotone{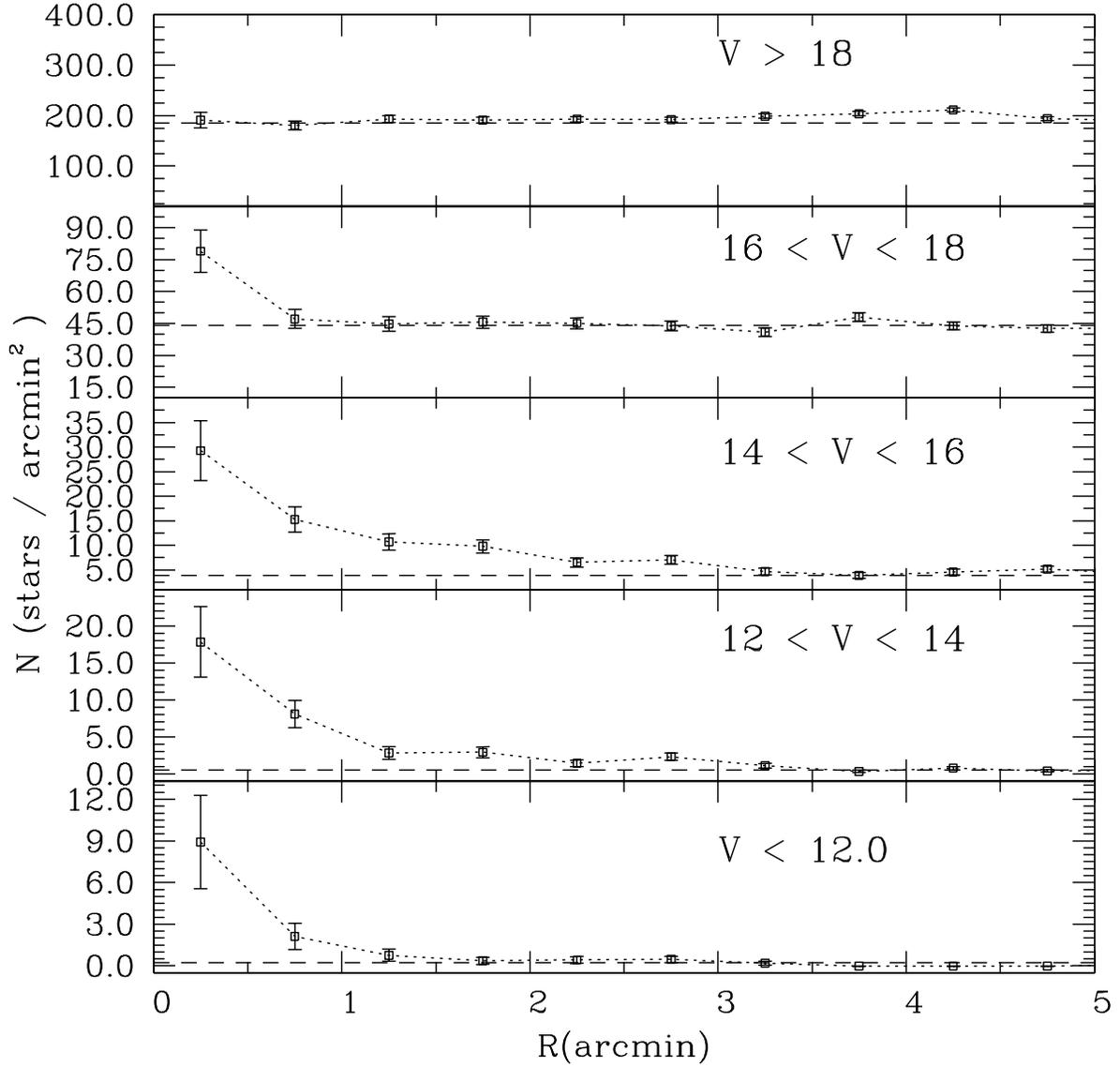}
\caption{Star counts as a function of radius from the adopted cluster
center for various magnitude intervals. The dashed line in each panel
indicates the mean density level of the surrounding Galactic disk
field in that magnitude level.}
\end{figure}

\begin{figure}
\plotone{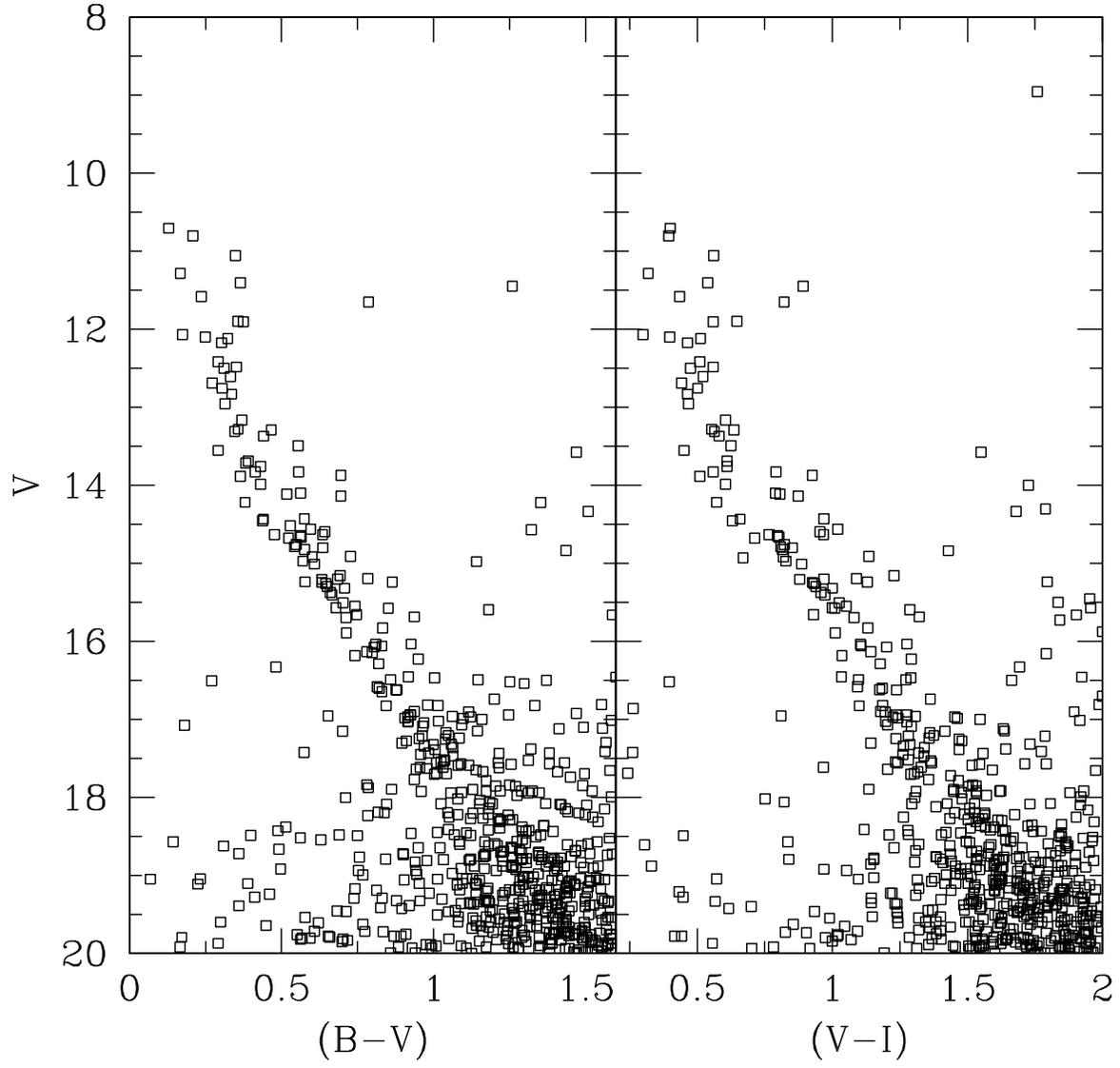}
\caption{CMDs for the stars in the field of NGC~6520.{\bf Left panel:}
V vs (B-V) CMD. {\bf Right panel:} V vs (V-I) CMD.}
\end{figure}

\begin{figure}
\plotone{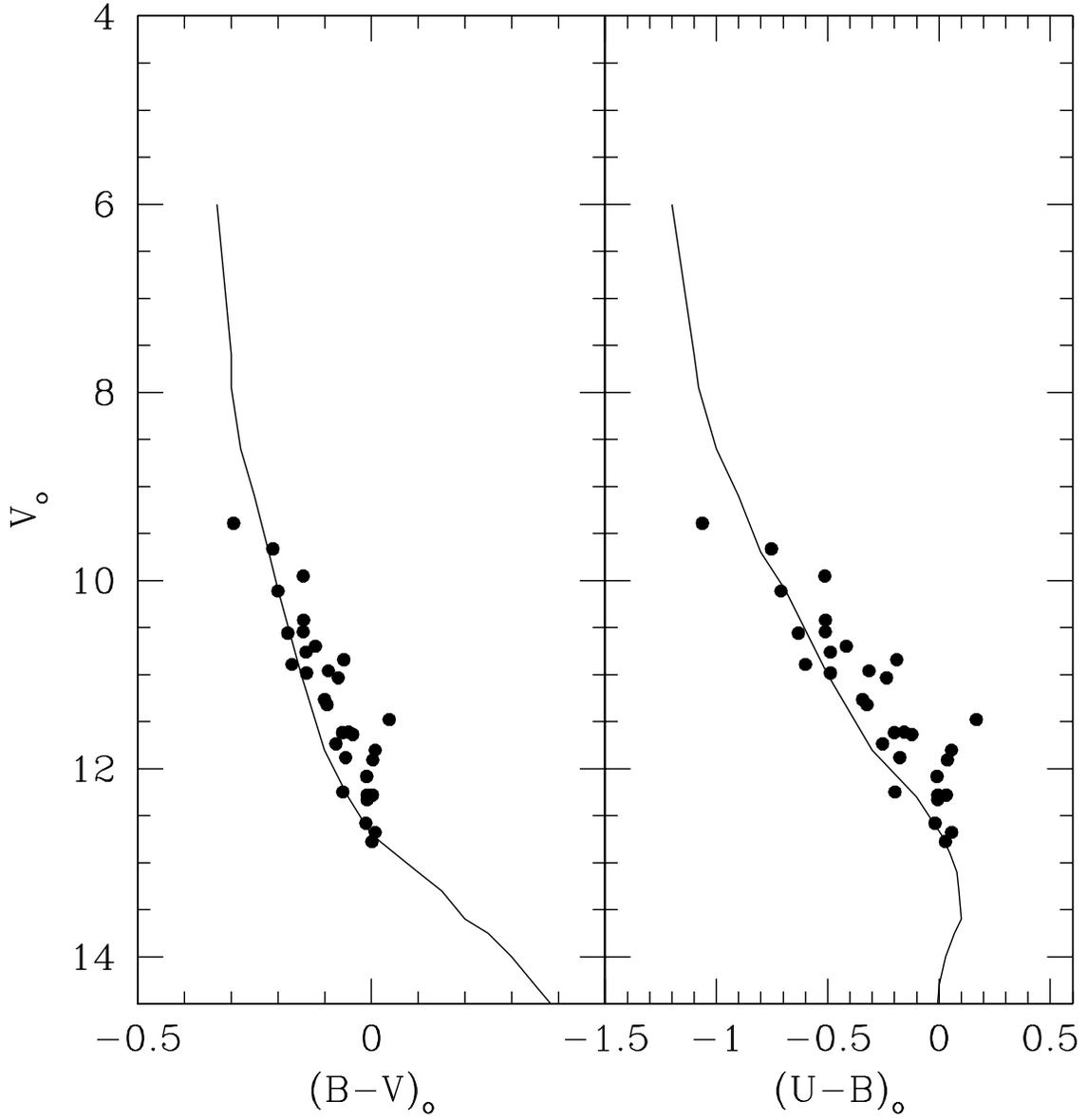}
\caption{
Reddening corrected CMDs from UBV photometry of \citet{kjf91}
Over-imposed in both panels is the empirical ZAMS from \citet{sch82}. See text for more details.}
\end{figure}

\begin{figure}
\plotone{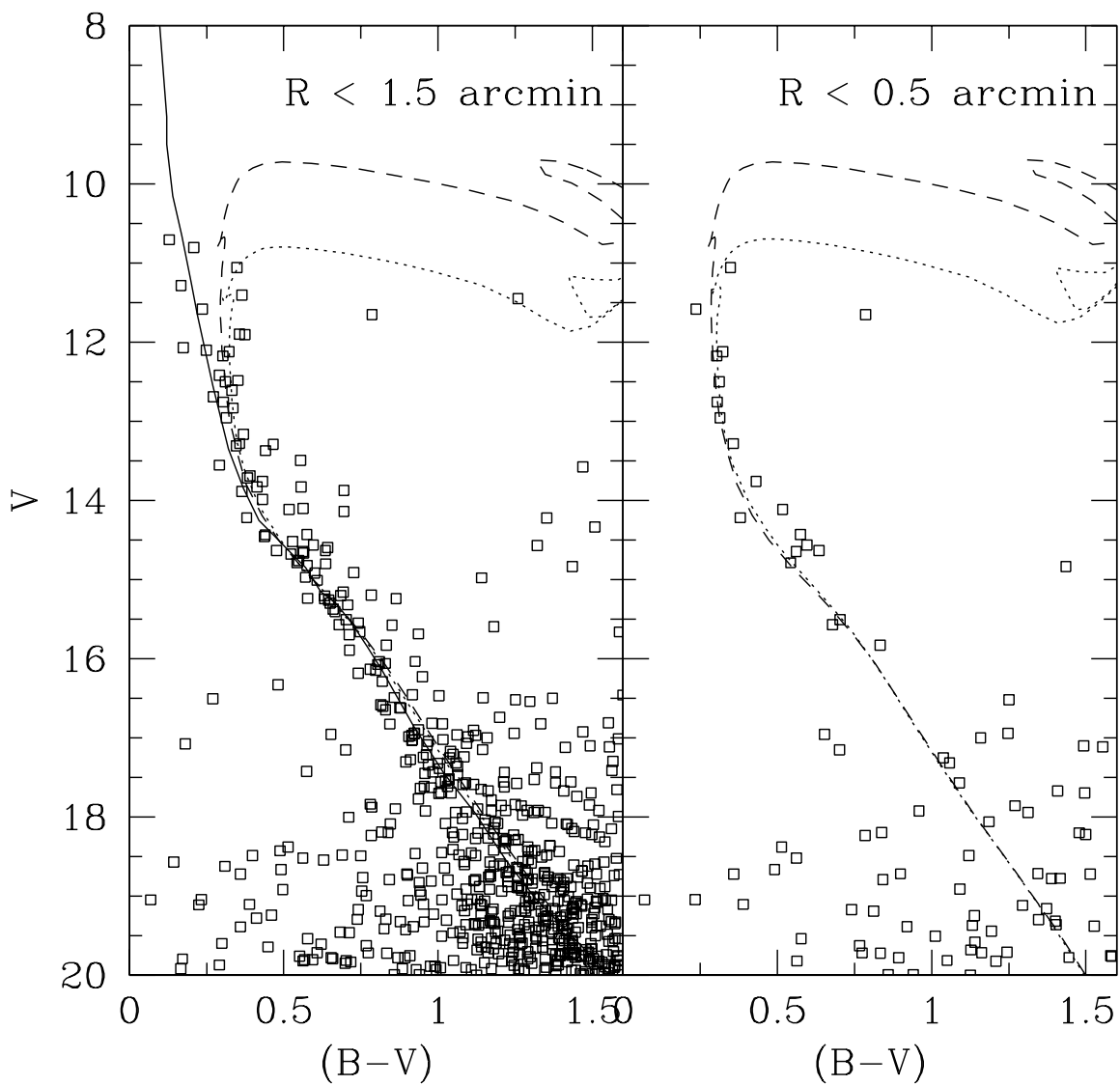}
\caption{CMDs of NGC~6520 as a function of the radius. Over-imposed are 
the empirical ZAMS from \citet{sch82} (solid line), and two solar-abundance isochrones
for the age of 100 (dashed line) and 200 (dotted line) million years. In the right panel
only star within 1 arcmin from the cluster center are considered to minimize the effect
of field star contamination in the age derivation.}
\end{figure}
\clearpage

\begin{figure}
\plotone{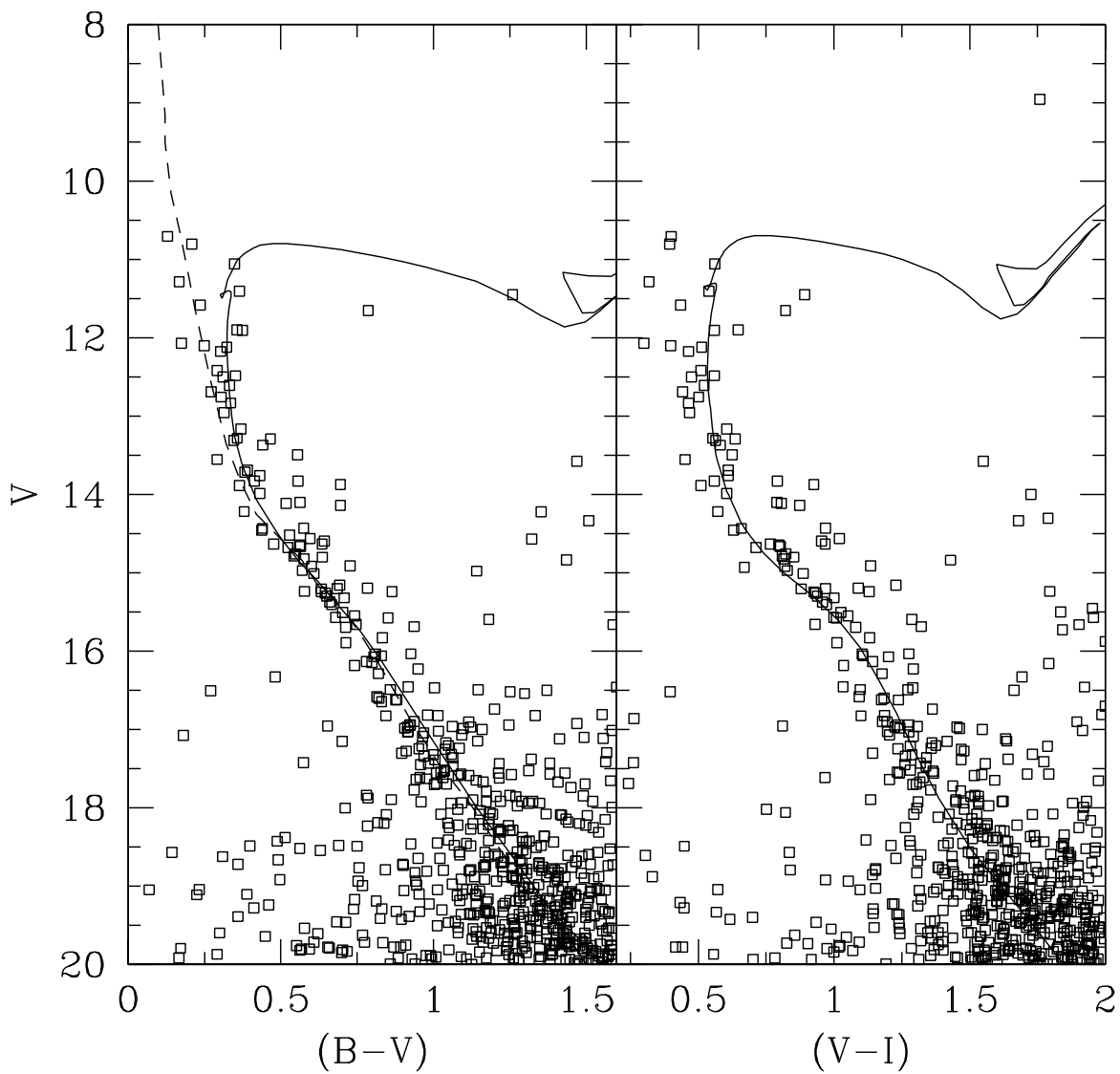}
\caption{The CMDs of Fig.~3. Over-imposed in the left panel is a \citet{sch82}
 ZAMS (dashed line)  and a solar metalliciy isochrone (solid line) 
for the age of 200 million years shifted by E$_{B-V}$=0.42 
and (m-M)$_V$=12.75(dashed line). In the right panel
the same isochrone is shifted by E$_{V-I}$=0.55 and (m-M)$_V$=12.75.}
\end{figure}
\clearpage

\begin{figure}
\plotone{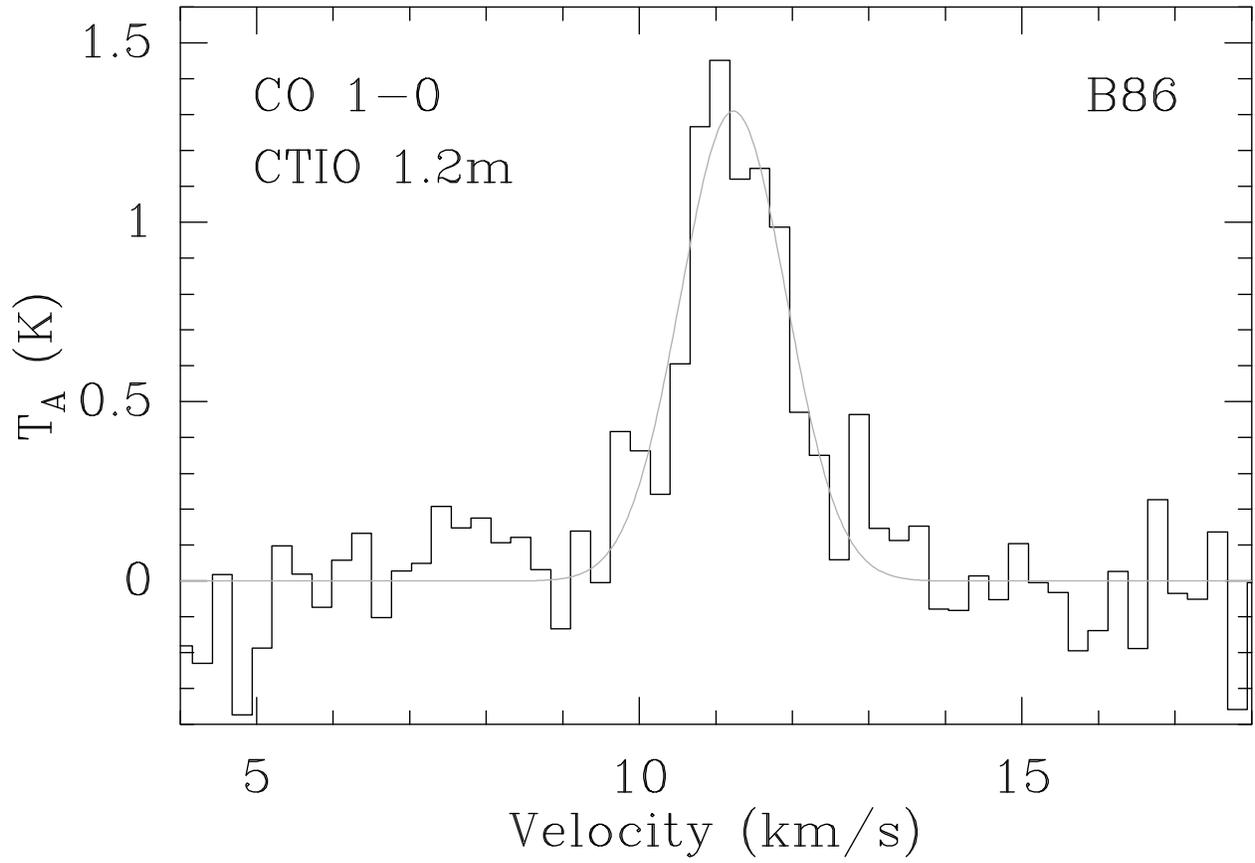}
\caption{Average of the two brightest CO 1-0 spectra in the region of Barnard~86.
The over-imposed line is the Gaussian fit.}
\end{figure}
\clearpage

\begin{figure}
\plotone{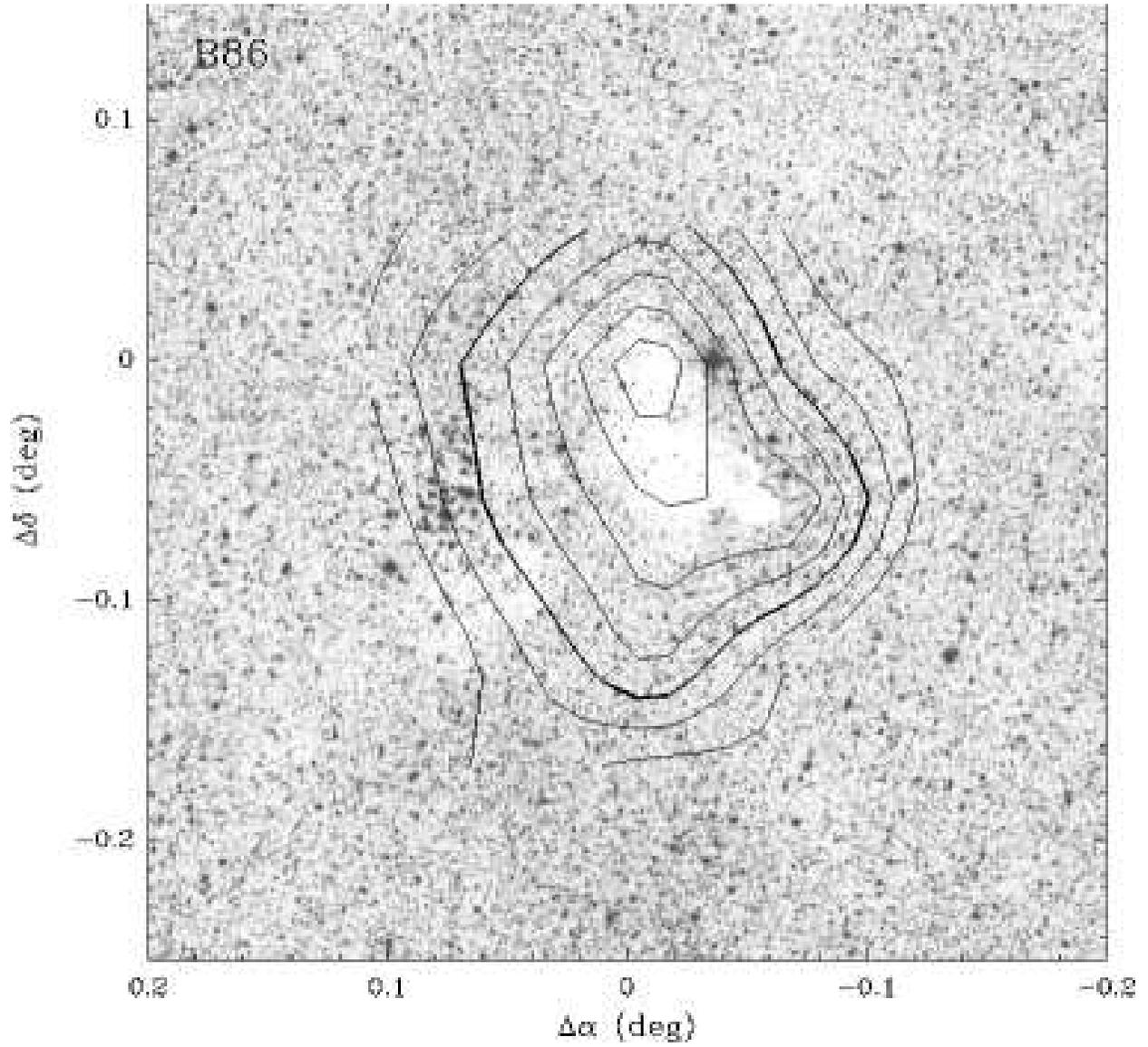}
\caption{DSS map of the region of NGC~6520 and Barnard~86. Over-imposed
are the iso-intensity contours of the radio observations.}
\end{figure}
\clearpage

\end{document}